\documentclass[aps, prl, reprint]{revtex4-1}
\usepackage{blindtext}
\usepackage{amsmath}
\usepackage{amssymb}
\usepackage{graphicx}
\usepackage{amsfonts}
\usepackage{lineno}
\usepackage{color}
\usepackage{multirow}
\usepackage{subfigure}
\usepackage{float}

\UseRawInputEncoding

\begin{document}
	\title{Influence of database  noises  to  machine learning for spatiotemporal chaos}

    \author{Yu Yang$^1$}
	\author{Shijie Qin$^1$}
	\author{Shijun Liao$^{1,2}$} \email{email address: sjliao@sjtu.edu.cn}

	\affiliation{$^1$ Center of Advanced Computing, School of Naval Architecture, Ocean and Civil Engineering, Shanghai Jiaotong University, China},
	
	\affiliation{$^2$ School of Physics and Astronomy,  Shanghai Jiaotong University, China}

	\begin{abstract}		
		A new strategy, namely the ``clean numerical simulation'' (CNS), was proposed  (J. Computational Physics, 418:109629, 2020) to gain reliable/convergent simulations (with negligible numerical noises) of spatiotemporal chaotic systems in a long enough interval of time, which  provide us benchmark solution for comparison.  Here we illustrate that machine learning (ML) can always  give good enough  fitting  predictions of a spatiotemporal chaos by using, separately, two quite different training sets:  one is  the ``clean database''  given by the CNS with negligible numerical noises,  the other is  the ``polluted database''  given by the traditional algorithms in single/double precision with considerably large numerical noises.   However, even in statistics, the ML predictions based on  the ``polluted  database''   are quite different from those based on the ``clean database''.  It illustrates that the database noises  have huge  influences on ML predictions of some spatiotemporal chaos, even in statistics.  Thus,  we must use a  ``clean''  database for machine learning of  some spatiotemporal chaos.   This surprising result might open a new  door and possibility to study machine learning.      			
	\end{abstract}

	\maketitle
		
    It was  first  discovered by Poincar\'{e} \cite{Poincare1890} that noises (or uncertainty) increase exponentially for chaotic dynamic systems due to the sensitivity dependence on initial conditions (SDIC).  The same phenomena was rediscovered by Lorenz \cite{Lorenz1963} in 1963 with a more popular name ``butterfly-effect'':  for a chaotic system, a tiny variation of initial condition might lead to a huge difference of numerical simulations after a long enough time \cite{sprott2010, Sprott2003Chaos}.   Furthermore,  Lorenz \cite{Lorenz2006Computational} discovered that computer-generated simulations of a chaotic system given by traditional methods (such as  Runge-Kutta method and so on) in single/double precision are sensitive not only to initial conditions but also to numerical algorithms.

	To overcome the above-mentioned limitations of traditional algorithms for chaotic systems,  Liao \cite{Liao2009, Liao2013A} proposed a new numerical strategy, called the ``Clean Numerical Simulation'' (CNS).   The CNS can greatly  reduce the global numerical noises to  a required tiny level so that a reliable (replicable) simulation can be obtained  in the whole spatial domain within a controllable interval of time $0 \leqslant t \leqslant T_c$, where $T_c$ is called ``critical predictable time''.  
Thus, the CNS can provide us a  reliable, true simulation of chaos  in a long enough interval of time, which can be used as a {\em benchmark} solution.  
    The CNS has been successfully applied to gain reliable computer-generated simulations of many chaotic systems, such as Lorenz system \cite{Liao2009, LIAO2014On}, three-body systems \cite{Liao2013B, Li2017More, Li2018Over, Li2018Collisionless}, some spatiotemporal chaotic systems  \cite{hu2020risks,  qin2020influence, Xu2021PoF}, two-dimensional Rayleigh-B\'{e}nard turbulence \cite{Lin2017On},  and so on.   All of these illustrate the validity of the CNS.  
    
Currently,  machine learning (ML) \cite{jaeger2001echo,  haynes2015reservoir, larger2017high, jaeger2004harnessing,   2017Using, lu2018attractor, 2018Model,2018Hybrid, 2019Model, tanaka2019recent, nakai2018machine, vlachas2020backpropagation} has been successfully applied in predicting evolution of many nonlinear dynamic system. Particularly, the reservoir computing (RC) \cite{jaeger2004harnessing, 2017Using, lu2018attractor, nakai2018machine, 2018Model, 2018Hybrid, 2019Model, tanaka2019recent, vlachas2020backpropagation} has shown significant success in modelling chaotic systems, which can alleviate the difficulty in learning recurrent connections of recurrent neural networks (RNNs) and besides decrease training cost.   
	
	 However,  all computer-generated chaotic simulations, which were used as database to evaluate performance of machine learning before, were given by the traditional algorithms in single/double precision, so that they  contain  considerably large numerical noises and thus unavoidably have great deviations from their corresponding ``true''  trajectories,  due to the butter-effect of chaos.   In other words, these  databases    contain  large artificial {\em noises}: they are ``polluted database'' for  machine learning.    
On the other side,  using the CNS, one can gain a true/reliable simulation of chaos in a long enough interval of time, which as a benchmark solution can provide us a ``clean database'' for machine learning.    These two computer-generated chaotic simulations of the same equation under the same initial/boundary conditions provide us two different training sets for machine learning.  Obviously, we can use the ``clean database'' as a benchmark  to  investigate whether or not the ML predictions  based on the ``polluted database''  have huge differences from those based on the ``clean database''.   This can reveal the influence of database noise to machine learning, which has been almost neglected by the ML community.  
       	
Without loss of generality,  let us consider here the Kuramoto-Sivashinsky (KS) equation
	\begin{align}
		& U_t+UU_x+U_{xx}+U_{xxxx}=0, \hspace{1.0cm} 0\leq x\leq L,    \label{KS}
	\end{align}
	which is a prototypical model of spatiotemporal chaos.   Like Khellat and Vasegh \cite{khellat2014kuramoto}, we choose the initial condition
	 \begin{eqnarray}
	  U(x,0) &= &-\sin\left(\alpha x \right)+\cos\left(\alpha x \right) \nonumber\\
	             &+& \sin\left(2 \alpha x \right)-\cos\left(2 \alpha x\right)    \label{Ini}
	 \end{eqnarray}
	 and the periodic boundary condition $U(0,t)=U(L, t)$, where $\alpha = 2 \pi/L$.   Following Lu {\em et al.}  \cite{lu2017data} and Lin \& Lu \cite{lin2021data}, we choose here  $L=21.55$,  which leads to $3$ linearly unstable modes with the maximum Lyapunov exponent $ \Lambda_{max}\approx0.04$,  corresponding to the Lyapunov time $T_L = 1/\Lambda_{max}\approx25$.
	 	
An effective  CNS algorithm for spatiotemporal chaos was 	 proposed by Hu \& Liao \cite{hu2020risks}.  First of all, $U(x,t)$ is discretized at $N$ equidistant points, say, $x_{k}= k \Delta x$,  where $k=0, 1, 2, ..., N-1$ and $\Delta x ={L}/{N}$.  Thus, $U(x,t)$ is approximated by a set of $N$ discretized time series 
	\[
	\left\{ U(x_{0},t), U(x_{1},t),U(x_{2},t),U(x_{3},t),\cdots,U(x_{N-1},t)  \right\}.
	\]
	The key point of the CNS \cite{hu2020risks, qin2020influence, Xu2021PoF} is to  reduce the global numerical noises (i.e. both of  truncation error and round-off error) so greatly that,  in a long enough interval of time, these noises are negligible, i.e. much smaller than $U(x,t)$ here.   
In the frame of the CNS,   the high-order Taylor expansion method
	\begin{equation}
		U(x_{k},t+\Delta t)=U(x_{k},t)+\sum^{M}_{m=1}U^{[m]}(x_{k},t)(\Delta t)^{m}    \label{Taylor}
	\end{equation}
	is used  to reduce the truncation error in the temporal dimension, where $M$ is the order of Taylor expansion, $\Delta t$ is time-step, with the definition 
	\begin{equation*}
		U^{[m]}(x_{k},t)=\frac{1}{m!}\frac{\partial^{m} U(x_{k},t)}{\partial t^{m}}.   \label{HB}
	\end{equation*}
	Obviously, the temporal  truncation error can be reduced to a required tiny level as long as $M$ is large enough and $\Delta t$ is reasonably small.   Differentiating $(m-1)$ times on both sides of Eq.~(\ref{KS})  with respect to $t$ and then dividing them by $m!$, we have 
	\begin{eqnarray}
			U^{[m]}(x_k,t) &=& -\frac{1}{m} \left[  \sum^{m-1}_{j=0}U^{[j]}(x_k,t)U^{[m-1-j]}_{x}(x_k,t) \right. \nonumber \\
			&+&	\left.  U^{[m-1]}_{xx}(x_k,t)+U^{[m-1]}_{xxxx}(x_k,t)\,\right],  \label{Taylor_G}
	\end{eqnarray}
	where $m\geq 1$, $U^{[0]}(x_{k},t)=U(x_{k},t)$ and 
\[
U^{[j]}_{x}(x_k,t) = \left. \frac{\partial U^{[j]}}{\partial x}\right|_{x=x_{k}}, U^{[j]}_{xx}(x_k,t) = \left. \frac{\partial^{2} U^{[j]}}{\partial x^{2}}\right|_{x=x_{k}},
\]	
and so on.  In the frame of the CNS,  
the spatial derivatives, such as $U^{[j]}_{x}(x_k,t)$, $U^{[j]}_{xx}(x_k,t)$ and $U^{[j]}_{xxxx}(x_k,t)$  in Eq.~(\ref{Taylor_G}), are calculated in rather high accuracy by means of the Fourier  series  of $U(x,t)$ in space, so as to  reduce the spatial truncation error.    
Here, the fast Fourier transform (FFT) algorithm is used.  Thus,       we can obtain an accurate enough approximation of the spatial derivatives as long as the mode-number $N$ of the spatial Fourier  series  is large enough. In this way, the spatial truncation error can be reduced to a required tiny level, too.   For details, please refer to Hu \& Liao \cite{hu2020risks}.

In addition,  unlike traditional algorithms that mostly  use single/double precision,  the multiple-precision \cite{oyanarte1990mp} with a large number of significant digits is employed in the frame of the CNS for {\em all} physical/numerical variables and parameters so as to reduce the round-off error to a required tiny level.  
In practice, in order to gain simulations more efficiently, the variable step-size (VS) scheme \cite{Barrio2005} is applied in the temporal dimension with a given allowed tolerance $tol$ of the governing equations.  Besides,  since the round-off error should be in the same level of the (temporal) truncation error, we always set the allowed tolerance $tol = 10^{-N_{s}}$, where $N_s$ is the number of significant digits  in multiple precision \cite{oyanarte1990mp}.
  In this way, we can {\em globally} control the spatial and temporal truncation errors by choosing a large enough mode-number  $N$ of the spatial Fourier series and a large enough number $N_s$ of significant digits for multiple precision \cite{oyanarte1990mp}, which corresponds to a small enough temporal allowed tolerance $tol=10^{-N_s}$ of the governing equation by means of the VS scheme \cite{Barrio2005}.  For details, please refer to Hu \& Liao \cite{hu2020risks}.  
	
		\begin{figure}[t]
		\begin{center}
			\begin{tabular}{c}
				\includegraphics[width=3.6in]{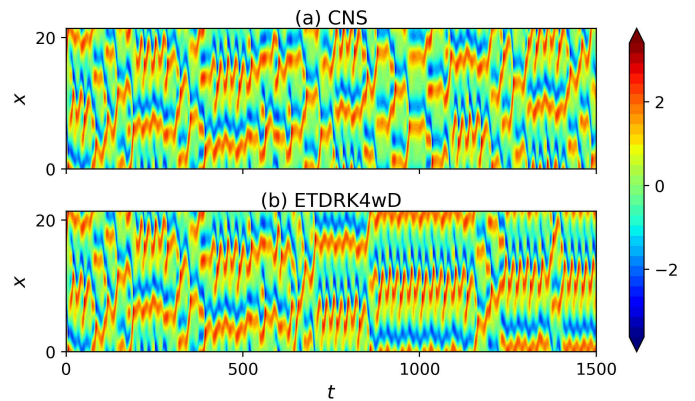}
			
			\end{tabular}
			\caption{Spatio-temporal trajectories of $U(x,t)$ of the KS equation, given by the CNS (a) and the traditional algorithm  ETDRK4wD (b), respectively.} \label{Contour}
		\end{center}\vspace{-0.6cm} 
	\end{figure}

	\begin{figure}[t]
	\begin{center}
		
		\includegraphics[width=2.5in]{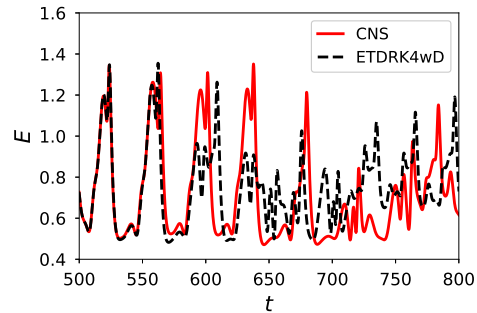}
		
		\caption{Comparison of the total energy $E(t)=\frac{1}{2L}\int^{L}_{0}U^2(x,t)dx$ given by the CNS and the  ETDRK4wD.}
		\label{etcompare}\vspace{-0.6cm} 
	\end{center}
\end{figure}

Furthermore,  to guarantee the reliability of a computer-generated simulation given by the CNS, an additional simulation with even smaller numerical noises is required to determine the so-called ``critical predictable time'' $T_{c}$  by means of comparing them with each other, so that both of them have no distinct differences in  $0\leq t \leq T_{c}$ within the whole spatial domain, as illustrated in \cite{hu2020risks}. 
	Solving the KS equation (\ref{KS}) by the CNS using different values of the mode-number  $N$ and the significant digit number $N_{s}$ for the multiple precision \cite{oyanarte1990mp}, we gained the linear relationship
	\begin{equation}
		T_c \approx \min \left\{ 18.6 N-387, ~46.4 N_s-222 \right\}.  \label{Tc_(N,Ns)}
	\end{equation}
	Thus, for a given value of  $T_{c}$, we can always find the corresponding mode-number $N$ of the spatial Fourier expansion and the significant digit number $N_s$ for multiple precision so as to gain a reliable chaotic numerical simulation of Eq.~(\ref{KS}) in $0\leq t \leq T_{c}$ within the whole spatial domain $x\in[0, L]$.  
Therefore,  according to (\ref{Tc_(N,Ns)}), we have $T_{c} \approx 4186$ in the case of  $N=256$, $N_s=95$ and $tol=10^{-95}$, which provides us a reliable  simulation (marked as CNS) of  the KS equation (\ref{KS}) in $0\leq t \leq 4000$.  It also provides us  the  ``clean database'' for the machine learning, which is used here as a benchmark to investigate the influence of database noises.

On the other side, 	following \cite{lin2021data},   we solved the KS equation (\ref{KS}) with the same initial condition (\ref{Ini}) and the same physical parameters in $0\leq t\leq4000$ by means of the 4th-order exponential time-differencing Runge-Kutta method \cite{cox2002exponential, kassam2005fourth} in double precision with the mode-number $N=108$ of the spatial Fourier expansion and time-step $\Delta t=10^{-3}$.   This provides us another numerical simulation (marked as ETDRK4wD)  with considerably  large numerical noises, i.e. the  ``polluted database''.    

Note that the ETDRK4wD simulation is given by the Runge-Kutta method in double precision (corresponding to $N_{s} = 16$) using the mode number $N = 108$, which are much smaller than the significant digit number $N_{s} = 95$ and the mode number $N = 256$ for the CNS simulation, respectively.  So,  the numerical noises of the ETDRK4wD simulation should be much larger than that of the benchmark solution  given by the CNS.  Thus,  due to the butterfly-effect of chaos,  the ETDRK4wD simulation should quickly become unreliable.   This is indeed true.  
As shown in Fig.~\ref{Contour} and Fig.~\ref{etcompare}, the trajectories of $U(x,t)$ and the total energy $E(t)=\frac{1}{2L}\int_0^L U^2(x,t) dx$ given by the ETDRK4wD  simulation agree with those of the benchmark solution (given by the CNS) only in a short  time $0\leq t\leq 560$:    the difference between the two simulations becomes distinct at $t=600$.  Thereafter, the numerical noises of the ETDRK4wD simulation  quickly increase to the {\em same} level of the ``true''  solution  so that it has obvious deviations from the benchmark result given by the CNS.

		\begin{figure}[t] 
		\begin{center}
			\includegraphics[width=3.5in]{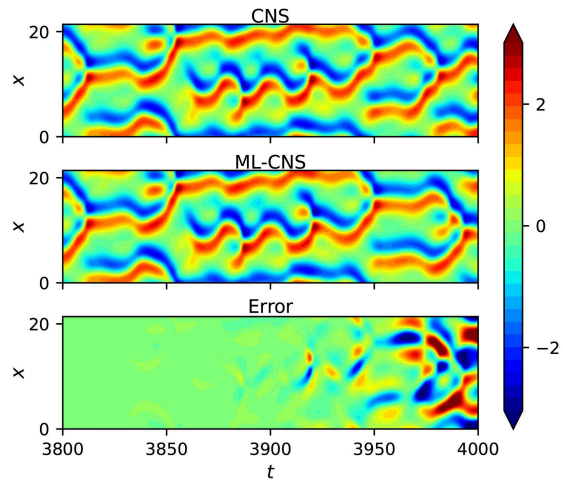}
		\caption{ Predictions of KS equation.  Top panel: actual data given by the CNS.  Middle panel:  ML prediction based on the ``clean database'' given by the CNS.  Bottom panel: error (middle panel minus top panel) in the reservoir predictions.}
			\label{precns}
		\end{center}\vspace{-0.6cm} 
	\end{figure}	
\begin{figure}
	\begin{center}

     	\includegraphics[width=3.5in]{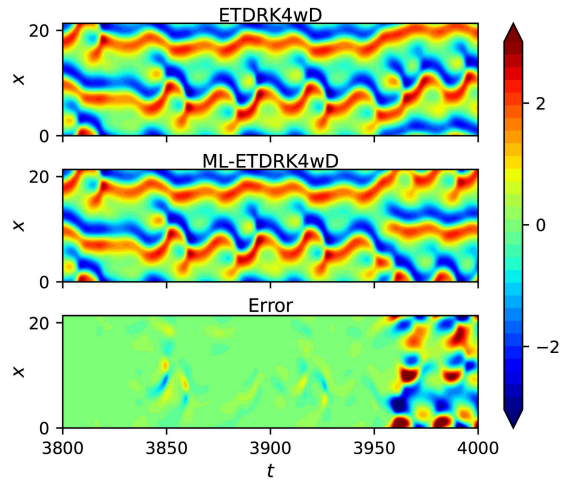}
		
		\caption{Predictions of KS equation.  Top panel: actual data given by the traditional algorithm ETDRK4wD. Middle panel:  ML prediction based on the ``polluted database'' given by the ETDRK4wD.  Bottom panel: error (middle panel minus top panel) in the reservoir predictions.}
		\label{predouble}
	\end{center}\vspace{-0.6cm} 
\end{figure}

	\begin{figure}[h]
		\begin{center}
		
		\subfigure[]{\includegraphics[width=2.4in]{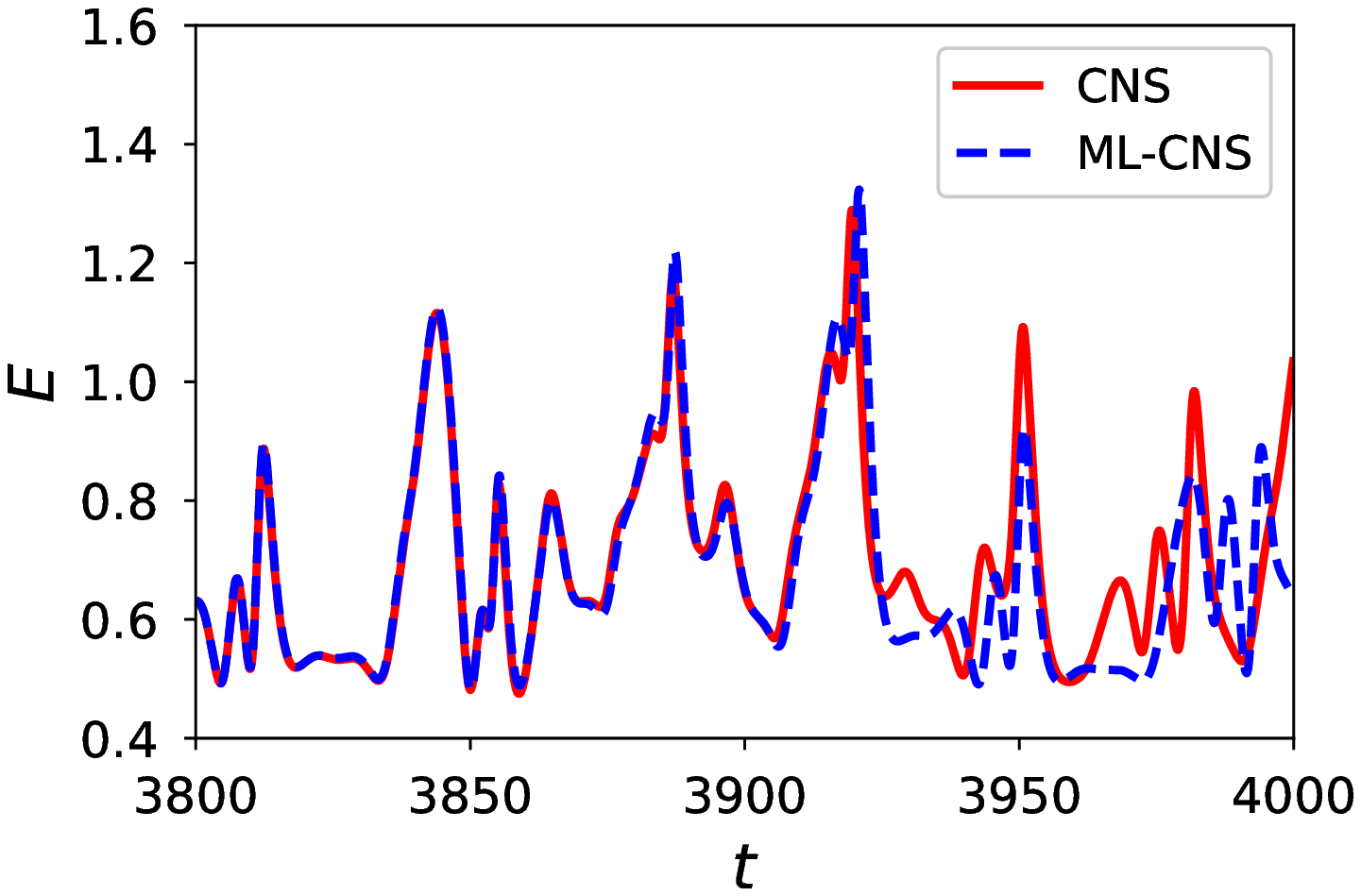}}
		\subfigure[]{\includegraphics[width=2.4in]{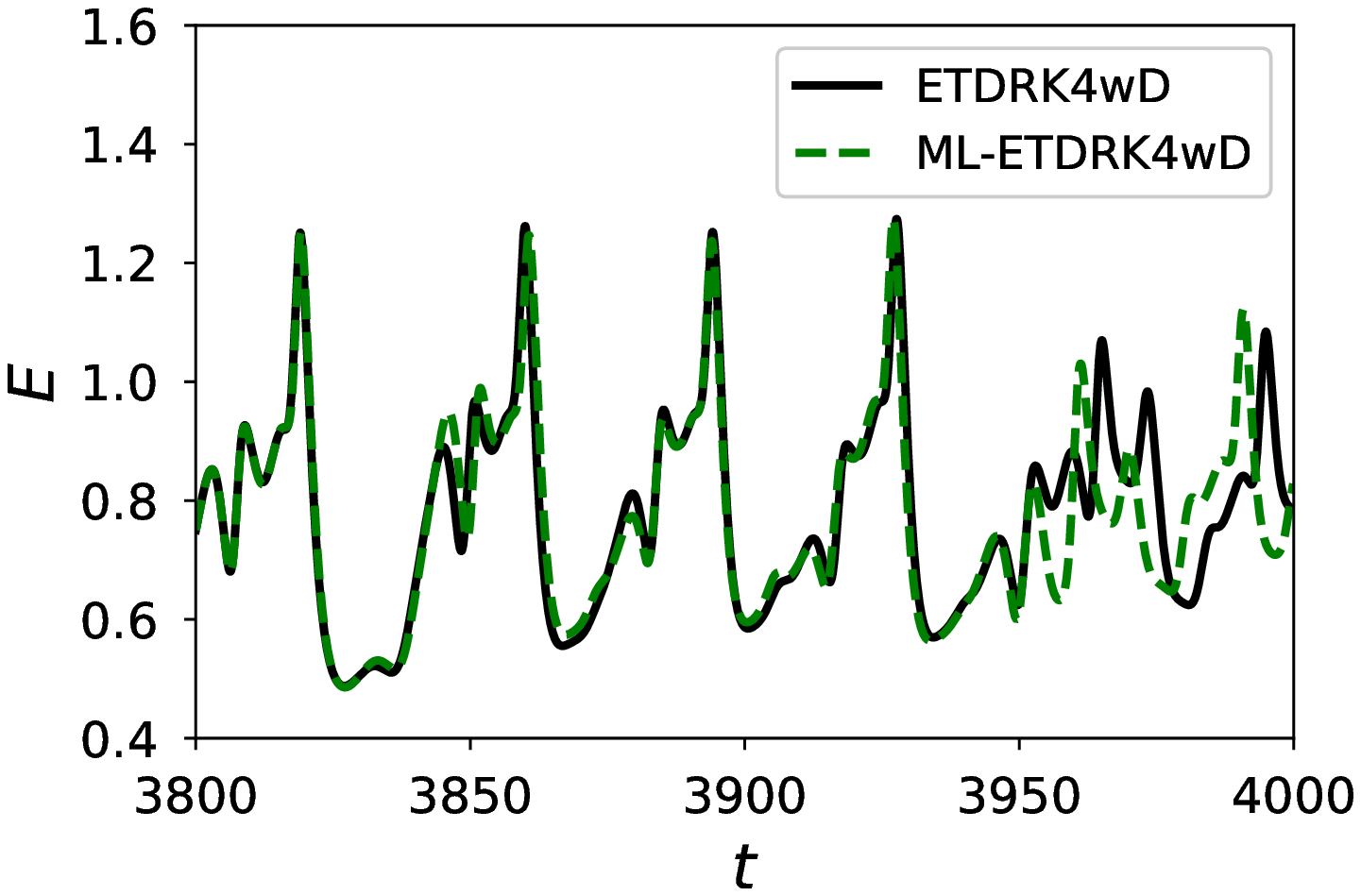}}
		
			\caption{Comparisons of the energy $E(t)=\frac{1}{2L}\int^{L}_{0}U^2(x,t)dx$. (a) Comparison between the actual result given by the CNS and the ML prediction  based on the ``clean database''. (b) Comparison between the actual result given by the traditional algorithm ETDRK4wD and the  ML prediction  based on the ``polluted database''.  Solid and dashed line denote the actual result and the ML prediction, respectively.}
			\label{ecns}
		\end{center}\vspace{-0.6cm} 
	\end{figure}

Following \citet{jaeger2001echo},  Jaeger \& Haas \cite{jaeger2004harnessing} and Pathak {\em et al.} \cite{2018Model},  we employ one widely used technique of machine learning (ML), i.e. the reservoir computing (RC),  to these two computer-generated simulations (of the KS equation in the same case) given by the ETDRK4wD and  CNS, respectively, corresponding to the ``polluted database''  and the ``clean database''.   
For the sake of simplicity, we designate the ML system based on  the ``clean database''  as ML-CNS, and the ML system based on the ``polluted database'' as ML-ETDRK4wD, respectively. 
We choose the observation time step $\Delta t = 0.1$, and use the results of the first 38000 time steps, i.e. $t\in(0, 3800)$,  as the training data.   The numbers of observation spatial grid of the training data  are 128 and 108 for the ML-CNS and ML-ETDRK4wD, respectively.  The basic ideas of the reservoir computing (RC)  are briefly described in the supplement, together with the corresponding hyper-parameters for the ML-CNS and ML-ETDRK4wD.

	\begin{figure}[t]
		\begin{center}
			
		\subfigure[]{\includegraphics[width=2.4in]{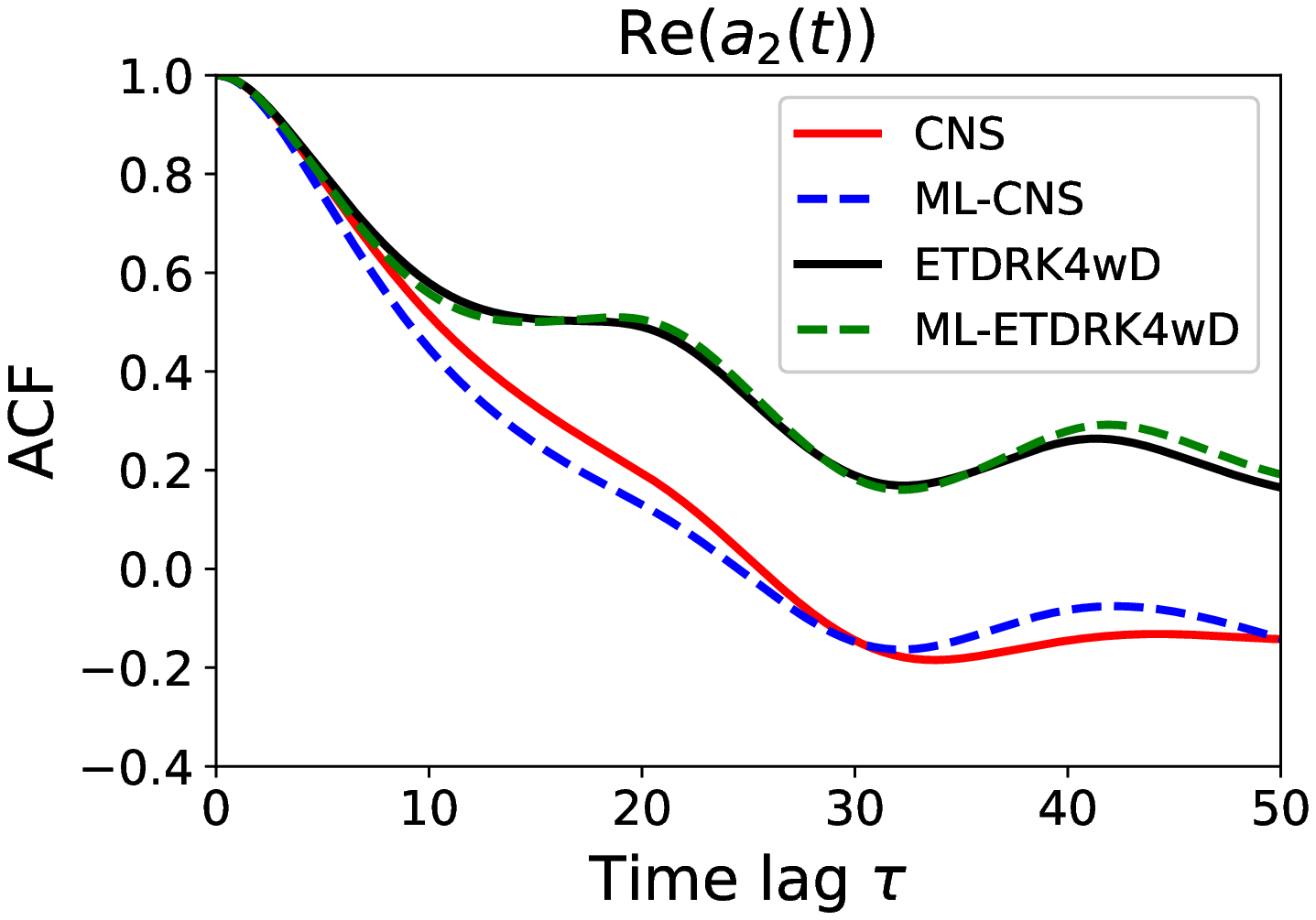}}
		\subfigure[]{\includegraphics[width=2.4in]{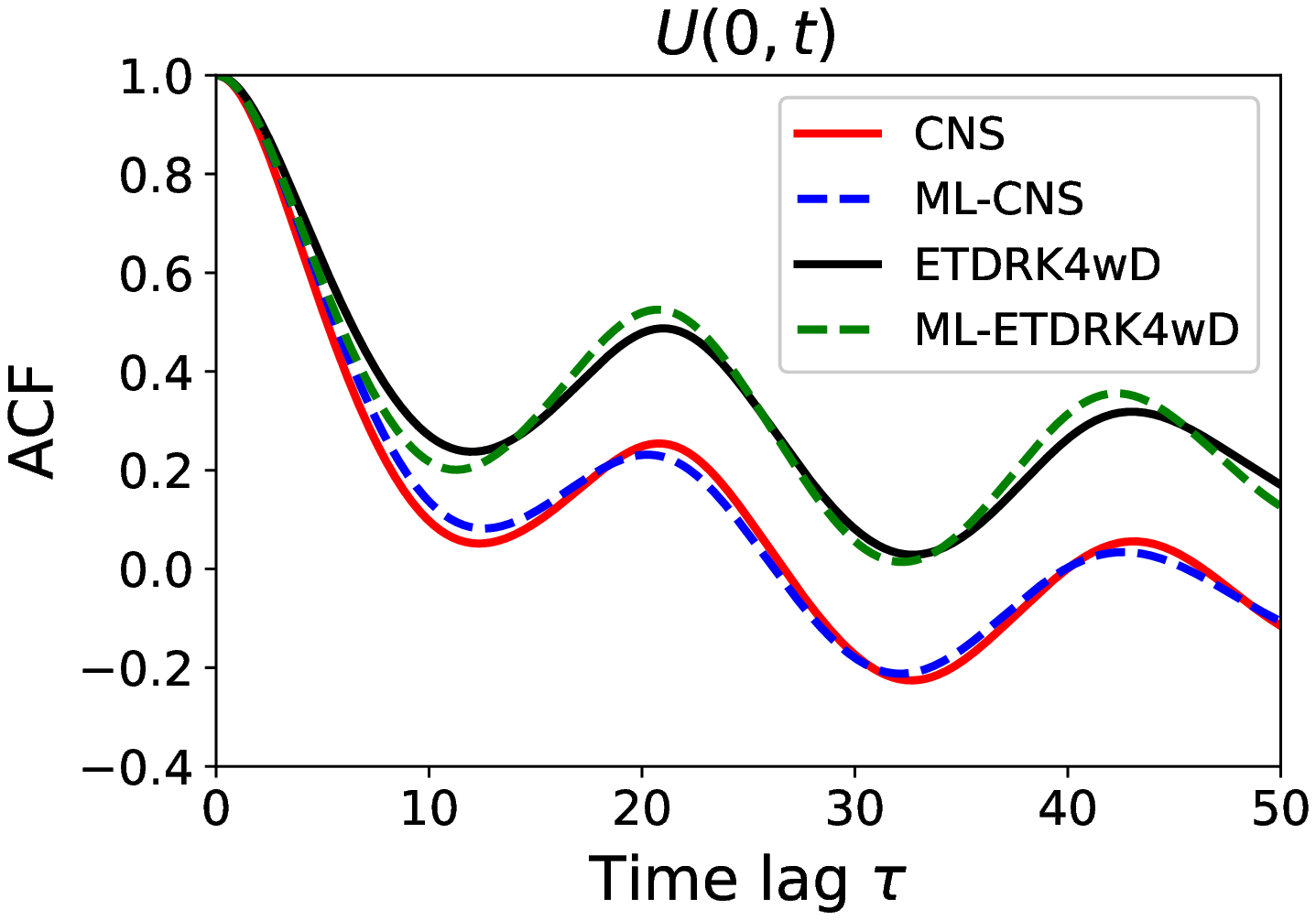}}
		
			\caption{Comparison of Autocorrelation functions (ACFs).  (a) ACFs of Re$(a_2(t))$. (b)  ACFS of  $U(0,t)$.  Solid line in red is the actual result given by the CNS,  the solid line in black is the actual result given by the traditional algorithm ETDRK4wD,  the dashed line in blue is the ML prediction  based on the ``clean database'', the dashed line in green is the  ML prediction  based on the ``polluted database'', respectively. }
			
		\label{acf}
		\end{center}\vspace{-0.6cm} 
	\end{figure}	
		
	Predictions are made respectively by these two ML systems in  $t\in[3800,7800]$ with  $\Delta t = 0.1$.   Fig.~\ref{precns} and Fig.~\ref{predouble} show the ML predictions (middle panel) of the spatiotemporal trajectories given by the ML-CNS (based on the ``clean database'') and ML-ETDRK4wD (based on the ``polluted database''), respectively, along with their  corresponding actual values (top panel) and deviations between the actual values and ML predictions (bottom panel).   In addition, Fig.~\ref{ecns} shows the comparisons of the total energy $E(t)$ predicted by the ML-CNS and ML-ETDRK4wD with their corresponding actual values.  
	The ML predictions of the spatiotemporal trajectories and the total energy given by the ML-CNS (based on the ``clean database'') agree well to their corresponding actual values in $t \in (3800, 3950)$ (about 6 Lyapunov time).    So do the ML predictions  by the ML-ETDRK4wD (based on the ``polluted database'') in $t \in (3800, 3950)$, too.   
	  However,  it must be emphasized that,  the  ML predictions of the spatiotemporal trajectories and the total energy based on the ``polluted database''  
 have the {\em distinct} deviations from those based on the ``clean database''.   It illustrates that data noises have a {\em great} influence on the ML predictions of the spatiotemporal trajectories and the  total energy.       

How about the statistic properties of the ML predictions?    Write
\[
U(x, t)= \sum^{N/2-1}_{n=-N/2}a_n(t) \; e^{\mathbf{i}\,nx},
\]
where $a_n(t)$ is the evolution of the Fourier mode.   As shown in Fig.~\ref{acf}(a), the autocorrelation function (ACF)  of the real part of $a_2(t)$ given by the ML-ETDRK4wD prediction   (based on the ``polluted database'')   in $t\in[3800,7800]$ agrees well with that by the actual ETDRK4wD  simulation.   Similarly, the ACF of the real  part of $a_2(t)$ given by the ML-CNS prediction (based on the ``clean database'') also agrees well with that by the actual CNS result.   For each database,  unlike the ML predictions of the spatiotemporal trajectories and the total energy that agree well with the corresponding actual values only in about {\em six} Lyapunov's times  $t \in [3800, 3950]$,  the ML predictions of the ACF (as a statistic result) are accurate enough  in a much {\em larger} interval of time $t\in[3800,7800]$.   In other words,  the {\em statistic} predictions  given by the machine learning for the spatiotemporal chaotic system are valid in a much longer interval of time  than the ML predictions of the spatiotemporal trajectories and the evolution of the total energy $E(t)$.        
However,  the  ACF of  the real part of $a_2(t)$ given by the ML  prediction based on the ``polluted database''  {\em significantly} deviates  from that by the ML prediction based on the ``clean database''.          

In a similar way  we can investigate the  ACFs of the ML prediction of $U(0, t)$.  It is found that the ACF of  $U(0, t)$ given by the ML prediction based on the ``polluted database''  also significantly deviates  from that by the ML prediction based on the ``clean database'',  as shown in Fig.~\ref{acf}(b).   These illustrate that the database noises might have huge influences on the ML predictions {\em even} in statistics.  

Machine learning (ML)  \cite{jaeger2001echo,   jaeger2004harnessing,  haynes2015reservoir, larger2017high, 2017Using, lu2018attractor, 2018Model,2018Hybrid, 2019Model, tanaka2019recent, nakai2018machine, vlachas2020backpropagation} is indeed a very promising tool that has many exciting applications in science and engineering.   However,   for machine learning of some spatiotemporal chaos,  database noises might lead to huge deviations from the true simulation, even in statistics.     Thus,  we {\em must} use a ``clean'' database for machine learning of  such kind of  spatiotemporal chaos, otherwise the ML predictions might lead to huge misunderstandings.   This surprising result might open a new  door and possibility to study machine learning.   Certainly, there are lots of work to do in future.

	~\\
	
	This work is partly supported by the National Natural Science Foundation of China (No. 91752104).

	\bibliography{ks_ref}
	\bibliographystyle{apsrev4-1} 

\end{document}